\documentclass[useAMS,usenatbib]{mn2e}

\usepackage{graphicx}
\usepackage{txfonts}
\usepackage{natbib}

\newcommand{\tacc}{\ensuremath{T_{acc}}}
\newcommand{\tdisk}{\ensuremath{T_{disk}}}
\newcommand{\tstar}{\ensuremath{T_\star}}

\newcommand{\tkh}{\ensuremath{T_{KH}}}

\newcommand{\rco}{\ensuremath{r_{co}}}
\newcommand{\ra}{\ensuremath{r_{A}}}
\newcommand{\rt}{\ensuremath{r_{t}}}

\newcommand{\wstar}{\ensuremath{\Omega_\star}}
\newcommand{\restar}{\ensuremath{\mathcal{R}_\star}}
\newcommand{\mstar}{\ensuremath{M_\star}}
\newcommand{\rstar}{\ensuremath{R_\star}}
\newcommand{\bstar}{\ensuremath{B_\star}}
\newcommand{\jstar}{\ensuremath{J_\star}}
\newcommand{\jmstar}{\ensuremath{j_\star}}
\newcommand{\lstar}{\ensuremath{L_\star}}
\newcommand{\mwind}{\ensuremath{\dot M^w_\star}}
\newcommand{\qstar}{\ensuremath{Q_\star}}

\newcommand{\msun}{\ensuremath{M_\odot}}
\newcommand{\rsun}{\ensuremath{R_\odot}}
\newcommand{\wsun}{\ensuremath{\Omega_\odot}}
\newcommand{\msunyr}{\ensuremath{M_\odot yr^{-1}}}
\newcommand{\mwsun}{\ensuremath{\dot M^w_\odot}}
\newcommand{\bsun}{\ensuremath{B_\odot}}

\newcommand{\jpl}{\ensuremath{J_{pl}}}
\newcommand{\jmpl}{\ensuremath{j_{pl}}}
\newcommand{\wpl}{\ensuremath{\Omega_{pl}}}
\newcommand{\mpl}{\ensuremath{M_{pl}}}
\newcommand{\rpl}{\ensuremath{R_{pl}}}
\newcommand{\bpl}{\ensuremath{B_{pl}}}
\newcommand{\taumig}{\ensuremath{\tau_{mig}}}
\newcommand{\acrit}{\ensuremath{a_{crit}}}

\newcommand{\wkep}{\ensuremath{\Omega_{Kep}}}
\newcommand{\macc}{\ensuremath{\dot M_{acc}}}

\newcommand{\tauaccup}{\ensuremath{\tau^{up}_{acc}}}

\title[Protostellar spin-down: a planetary lift?]{Protostellar spin-down: a planetary lift?}
\author[J. Bouvier, D. C\'ebron]{J. Bouvier$^{1,2}$\thanks{E-mail:
    Jerome.Bouvier@obs.ujf-grenoble.fr}, D. C\'ebron$^{3,4}$\thanks{E-mail:
    David.Cebron@ujf-grenoble.fr}\\
$^{1}$Universit\'e Grenoble Alpes, IPAG, F-38000 Grenoble, France\\
$^{2}$CNRS, IPAG, F-38000 Grenoble, France\\
$^{3}$Universit\'e Grenoble Alpes, ISTerre, F-38000 Grenoble, France\\
$^{4}$CNRS, ISTerre, F-38000 Grenoble, France
}

\begin{document}

\date{Accepted 2015 August 4.  Received 2015 August 3; in original form 2015 June 10}

\pagerange{\pageref{firstpage}--\pageref{lastpage}} \pubyear{2015}

\maketitle

\label{firstpage}

\begin{abstract}
When they first appear in the HR diagram, young stars rotate at a
mere 10\% of their break-up velocity. They must have lost most of the
angular momentum initially contained in the parental cloud, the
so-called angular momentum problem. We investigate here a new
mechanism by which large amounts of angular momentum might be shed
from young stellar systems, thus yielding slowly rotating young
stars. Assuming that planets promptly form in circumstellar disks and
rapidly migrate close to the central star, we investigate how the
tidal and magnetic interactions between the protostar, its close-in
planet(s), and the inner circumstellar disk can efficiently remove
angular momentum from the central object. We find that neither the
tidal torque nor the variety of magnetic torques acting between the
star and the embedded planet are able to counteract the spin up
torques due to accretion and contraction. Indeed, the former are
orders of magnitude weaker than the latter beyond the corotation
radius and are thus unable to prevent the young star from spinning
up. We conclude that star-planet interaction in the early phases of stellar evolution does not appear as a viable alternative to magnetic star-disk coupling to understand the origin of the low angular momentum content of young stars.
\end{abstract}


\section{Introduction}

It has long been known that the Sun and its siblings are characterized by quite modest rotational velocities, of order of a few km/s at most \cite[e.g.][]{Kraft70}. This stands in sharp contrast to more massive stars that exhibit much higher spin rates all over their evolution. \citet[][]{Schatzman62} first theorized that magnetized winds from solar-type stars would carry away a significant amount of angular momentum, thus braking the stars quite efficiently. Hence, regardless of their initial velocity, stars with outer convective envelopes would end up as slow rotators on the main sequence. This expectation was later confirmed by \citet[][]{Skumanich72} who showed that the rotational velocity of solar-type stars appeared to steadily decrease on the main sequence, following the well-known $\Omega \propto t^{-1/2}$ relationship.   

Extrapolating this relationship back in time to the early pre-main sequence (PMS), young stars were thus expected to be fast rotators at birth. More generally, as gravity dominates the late stages of protostellar gravitational collapse, newly-born stars were commonly thought to rotate close to break-up velocity.  Surprinsingly, the first measurements of spin rates for low-mass PMS stars, the so-called T Tauri stars (TTS), did not meet these expectations. On the contrary, young stars were found to have only moderate rotational velocities, on average about 10 times that of the Sun's, i.e., a mere 10\% of their break-up velocity \cite[][]{VogelKuhi81, Hartmann86, Bouvier86}. More than 30 years later, this aspect of the so-called "initial angular momentum problem" remains very much vivid. 

Several physical processes have been proposed to account for the slow rotation rates of young solar-type stars. They all rely on the magnetic interaction between the young star and its circustellar disk. At least 3 classes of models can be identified: X-winds \cite[][]{Shu94}, accretion-powered stellar winds \cite[][]{MattPudritz05}, and magnetospheric ejections \cite[][]{ZanniFerreira13}. All these models investigate how the angular momentum flux between the star and its surrounding is modified by the magnetospheric interaction with the inner accretion disk. A brief outline of these models and a discussion of their specific issues can be found in \citet[][]{Bouvier14}. While a combination of processes may eventually provide strong enough braking torques on young stars to account for their low spin rates, none have definitely proved to be efficient enough. 

We investigate here an alternative process to account for the low angular momentum content of young stars. Specifically, we explore the flux of angular momentum being exchanged within a system consisting of a magnetic young stars surrounded by a close-in orbiting planet embedded in the inner circumstellar disk. We review tidal and magnetic interactions between the protostar and a close-in planet to estimate whether spin angular momentum can be transfered from the central star to the planet's orbital momentum and eventually from there to the disk by gravitational interaction, thus effectively spinning down the central star.  Previous papers have extensively explored  star-disk, star-planet, and planet-disk interactions, usually with the aim to investigate the orbital evolution of close-in planets  \citep[e.g.][]{ZhangPenev14, Laine2008, Laine2012, Chang2010, Chang2012, Lai2012}. We build here from these earlier studies to explore the star-planet-disk interaction in the specific framework of the initial angular momentum problem. Indeed, while addressing the issue of halting planet migration near young stars through magnetic torques, \citet[][]{Fleck08} mentions that such interactions may be at least partly responsible for the slow rotation rates of PMS stars. We will show here  that, for realistic sets of stellar and planetary parameters, magnetic as well as tidal torques seem actually unable to prevent pre-main sequence stellar spin-up.  

In Section 2 we describe the general idea developed here, summarize the parameters of the system, and express the requirement for an effective protostellar spin down. In Section 3 we summarize the accelerating torques acting on the central star, namely accretion and contraction, that tend to spin it up. In Section 4, we explore both tidal torques and magnetic torques acting between the young star and the inner planet and review their efficiency in removing angular momentum from the central object. We discuss the quantitative torque estimates and the limits of the model in Section 5, and highlight our conclusions in Section 6.    

\section{Star -- Planet -- Inner Disk Interaction (SPIDI): a general
framework}

\begin{figure}
   \centering
   \includegraphics[width=9cm]{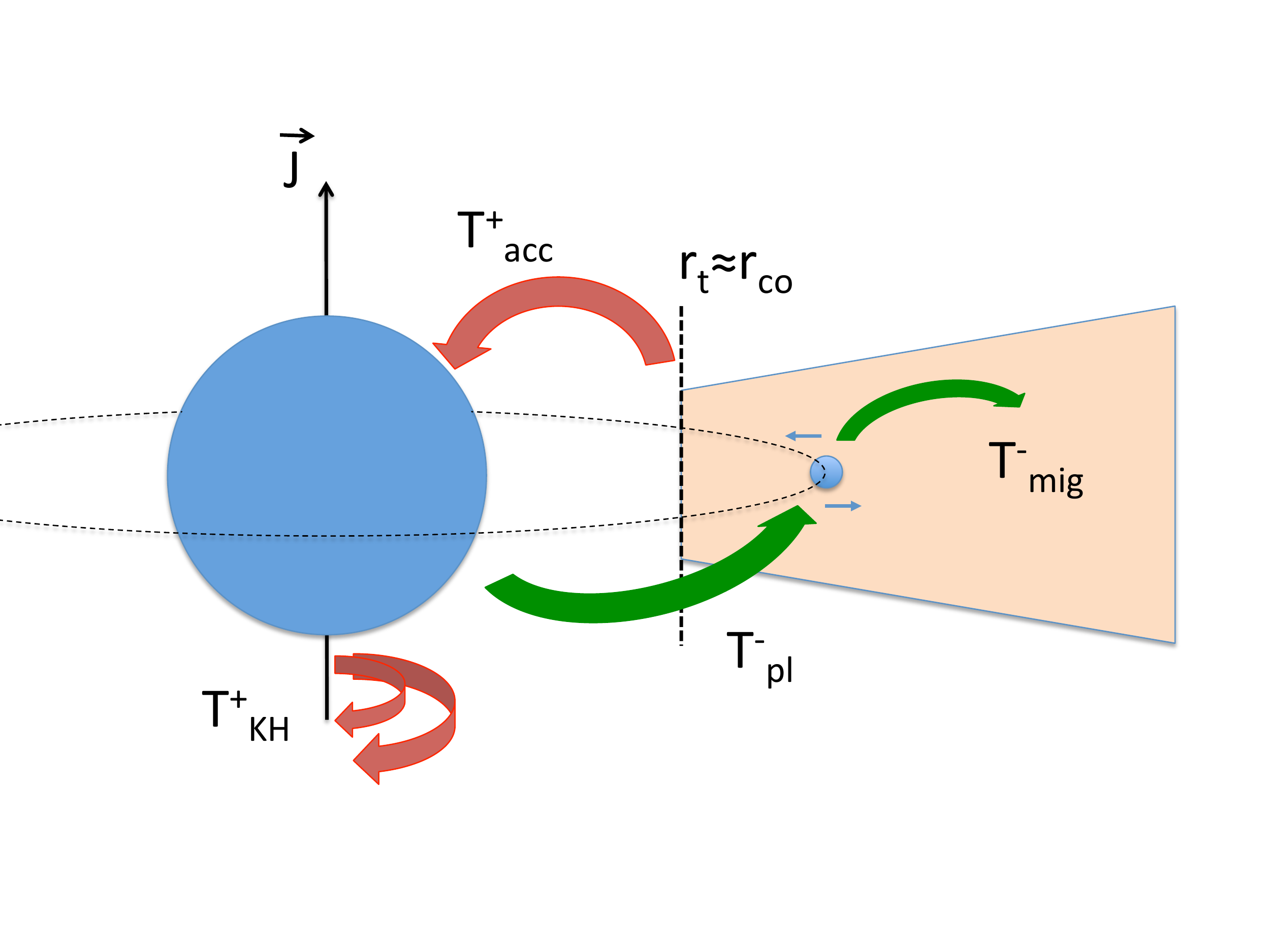}
      \caption{A sketch of the star-planet-inner disk interaction, with the main torques indicated: accretion, contraction, star to planet, planet to disk. 
              }
         \label{spidi}
   \end{figure}

The general idea that we develop here is to investigate whether a fraction of the spin angular momentum of the star can be transferred to the orbital angular momentum of a close-in planet, which in turn could lose its excess angular momentum to the disk through tidal interaction. In this way, the planet would act as a "lift" carrying angular momentum from the central star to the outer disk (cf. Fig.~\ref{spidi}). We aim here at computing an equilibrium solution that would fulfill this requirement, and hence prevent the star from spinning up as it accretes and contracts. Alternatively, a more dynamical view of this process can be thought of, where the planet migrates successively inwards and outwards several times just outside of the corotation radius, thus transferring over time a significant fraction of the star's angular momentum to the outer disk.  

In the following subsections we compute the torques acting in a system consisting of a young star, its circumstellar disk, and a close-in orbiting planet. The parameters of the system are listed in Table~\ref{param}. To set orders of magnitude, we compute the specific angular momentum of the star, \jmstar and of the planet, \jmpl. 

The specific spin angular momentum of the central star is
\begin{equation}\label{jmstar}
\jmstar = {{\jstar}\over{\mstar}} = k^2 \rstar^2 \wstar , 
\end{equation}
with k$^2$=0.205 for a completely convective star, which yields $j_\star$ = 5.8 10$^{16}$ cm$^2$s$^{-1}$. 
The specific orbital angular momentum of a non-eccentric planet orbiting at a distance $a$ from the star is:
\begin{equation}\label{jmpl}
\jmpl = {{\jpl}\over{\mpl}} = (G \mstar a)^{1/2}
\end{equation}
which is of order of a few 10$^{18}$  cm$^2$s$^{-1}$ for an orbital semi-major axis of a few stellar radii. The specific angular momentum of the planetary orbit is 2 orders of magnitude larger than the specific angular momentum contained in the stellar spin. However, a Jupiter-mass planet orbiting at the corotation radius has nearly the same total angular momentum as the star,
\begin{equation}\label{jspinorbit}
{{\jpl}\over{\jstar}} = {{\mpl (G\mstar a)^{1/2}}\over{k^2 \mstar \rstar^2 \wstar}} = {{\mpl}\over{k^2\mstar}}\left({{\Omega_{br}}\over{\wstar}}\right)^{4/3} \simeq 0.2 ,
\end{equation}
for $a = \rco = \wstar^{-2/3}(G\mstar)^{1/3}$, the break-up velocity $\Omega_{br}=(G\mstar)^{1/2}\rstar^{-3/2}$, and the system parameters defined in Table~\ref{param}. 

\section{Spin-up torques: young stars ought to rotate fast}

In this section, we describe torques that act to spin up young stars as they start their evolution on Hayashi tracks, namely the accretion and the contraction torques. 

\begin{table}
\caption{System parameters}             
\label{param}      
\centering                          
\begin{tabular}{l l }        
\hline\hline                 
Star: & \mstar = 1 \msun; \rstar = 2\rsun; \wstar = 5 \wsun;  $\bstar= 1 kG$\\
Disk/wind: & \macc = 10$^{-9}$\msunyr ; \mwind = 10$^{-10}$\msunyr \\ 
Planet: & \mpl = 10$^{-3}$ \msun; \rpl=0.2\rsun; \taumig = 1 Myr \\
\hline                        
\end{tabular}
\end{table}

\subsection{The accretion torque}

A magnetic young star accreting from its circumstellar disk assumed to be in Keplerian rotation undergoes a spin-up torque expressed as
 \begin{equation} \label{eq:tacc}
\tacc = \macc \rt^2 \wkep(\rt) = \macc (G \mstar \rt)^{1/2} ,
\end{equation}
where \macc\ is the mass accretion rate onto the star, \rt\ is the inner disk truncation radius, \wkep\ the Keplerian velocity in the disk, and \mstar\ the stellar mass. 

For the system parameters listed in Table~\ref{param}, the stellar magnetosphere truncates the disk at a distance of
\begin{equation}\label{eq:trunc}
\rt \approx 2\bstar^{4/7} \macc^{-2/7} \mstar^{-1/7} \rstar^{12/7} = 3.6\ 10^{11}\, \textrm{cm} = 5 \rsun = 2.5 \rstar ,
\end{equation}
where \bstar\ is the star's magnetic field and \rstar\ the stellar radius \cite[][]{Bessolaz08}. 

Alternatively, assuming that the truncation radius is located close to the corotation radius in the disk
\begin{equation}\label{eq:rco}
\rt \simeq \rco = \wstar^{-2/3}(G\mstar)^{1/3} =  12\rsun = 6 \rstar ,
\end{equation}
 where \wstar\ is the star's angular velocity. 

For these values of the truncation radius, the accretion torque
amounts to \tacc = 4.3-6.7 10$^{35}$ g cm$^2$ s$^{-2}$. The spin up timescale of the central star due to accretion is 
\begin{equation}\label{eq:tauacc}
\tauaccup = \jstar/\tacc = {{k^2 \mstar \rstar^2 \wstar}\over{\macc (G\mstar\rt)^{1/2}}} ,
\end{equation}
with the stellar angular momentum \jstar = 1.2 10$^{50}$ g cm$^2$ s$^{-1}$. Using the system parameters  listed in Table~\ref{param}, the spin up timescale thus amounts to 7 Myr and reduces to 0.7 Myr for a protostellar mass accretion rate of 10$^{-8}$\msunyr. Hence, as shown by \citet[][]{HartmannStauffer89}, accretion from the circumstellar disk is expected to spin up the star to a significant fraction of its break-up velocity within a few Myr.  

\subsection{The contraction torque}

As newly-born stars descend their Hayashi track, their radius decreases and they eventually develop a radiative core. Both effects yield a reduction of the stellar moment of inertia and, if angular momentum is conserved, the star spins up. Taking into account only radius contraction during the first few Myr, the fully convective star can be described as a $n=3/2$ polytrope. 

The potential energy of a polytrope of index $n$, is given by
\begin{equation}\label{eq:wpot}
E_{pot} = - {{3}\over{5-n}} {{G\mstar^2}\over{\rstar}}
\end{equation}
and the luminosity of a star undergoing homologous contraction is
\begin{equation}\label{eq:lum}
\lstar = - {{(3\gamma  - 4)}\over{3(\gamma-1)}} {{dE_{pot}}\over{dt}} = 4 \pi \rstar^2 \sigma T_{eff}^4 .
\end{equation}
The torque that should be applied to prevent the star from spinning up  as it contracts down on the pre-main sequence on a Kelvin-Helmholtz timescale is
\begin{equation}\label{eq:tkh}
\tkh = \wstar {{\textrm{d}I}\over{\textrm{d}t}} = 2 k^2 \mstar \wstar \rstar {{\textrm{d}\rstar}\over{\textrm{d}t}} ,
\end{equation}
and combining equations \ref{eq:wpot}-\ref{eq:tkh}, with $n=3/2$ and $\gamma=5/3$, one finally gets
\begin{equation}\label{eq:tkhlum}
\tkh = {{14}\over{3}} k^2 {{\wstar\rstar^3}\over{G\mstar}} \lstar .
\end{equation}
With the reference parameters listed in Table~\ref{param} and assuming T$_{eff}$ = 4000 K,  one finds \tkh $\simeq$ 5 10$^{35}$ g cm$^2$ s$^{-2}$. Hence, stellar contraction is equivalent to an accelerating torque  whose magnitude is similar to that of the accretion torque.  Both contraction and accretion will thus equally act to spin the young star up. 

\section{Spin down torques: can they prevent fast rotation?}

In this section, we investigate the torque a planet orbiting outside
the corotation radius would exert onto the central star. Such a planet
would extract angular momentum from star and migrate outwards. At the
same time, the planet feels the tidal torque from the disk through
Lindblad resonances and tends to migrate inwards. An equilibrium
configuration could result if the torques acting on the planet balance
outside the corotation radius, thus ensuring a continuous transfer of
angular momentum from the star to the planet and from the planet to
the disk (cf. Fig.~\ref{spidi}). If the outwards net flux of angular
momentum can counteract the accretion and contraction torques, the
star will thus be prevented from spinning up. Whether such an
equilibrium configuration can be reached, and whether it is able to
effectively spin the star down, depends on the nature and strength of
the star-planet-disk interaction. We first discuss tidal effects, and
then turn to magnetic interactions.

\subsection{Tidal torques}

We investigate whether tidal torques are strong enough to allow the orbiting planet to extract spin angular momentum from the star and to pass it on to the outer disk. In order to spin the star down, the planet must be located beyond the corotation radius. The inward migration torque the disk-embedded planet experience must therefore be balanced by the star-planet tidal torque beyond the corotation radius for the lift to operate. 

The migration torque exerted by the disk on the planet can be expressed as
\begin{equation}\label{eq:tdisk}
\tdisk = {{\textrm{d}\jpl}\over{\textrm{d}t}} = {1\over2} \mpl (G\mstar)^{1/2} a^{-1/2} {da\over dt} \sim {\jpl\over\taumig} ,
\end{equation}
where \taumig\ is the planet's migration timescale in the inner disk. 

We may first verify that the magnitude of the migration torque outside the corotation radius would be able to compensate for the accretion and contraction torques. The condition can be written as
\begin{equation}\label{cdttidal}
\left|\tdisk(a\geq\rco)\right| \geq \left|\tacc + \tkh\right| \simeq 2\left|\tacc\right| .
\end{equation}
Using the above expressions for the accretion and migration torques, and assuming the inner disk is located at the corotation radius (cf. Eq.~\ref{eq:rco}), this translates to
\begin{equation}\label{accmig}
{a\over\rco} \geq \left( 2 \taumig {\macc\over\mpl}\right)^2 
\end{equation}
which, adopting parameters from Table~\ref{param}, yields
$a\geq4\rco$. {\it Hence, if a giant planet can be hold off beyond
  the corotation radius, it may effectively extract enough angular
  momentum so as to compensate for the accretion and contraction
  torques, 
  thus preventing the central star from spinning up.}

Is the tidal effect between the star and the planet strong enough to prevent the planet from migrating inwards of the corotation radius? 
The tidal torque is given by\footnote{The factor of 2 difference between Eq.~\ref{eq:tstar} above and Eq.(18) of \citet{Chang2010} arises from their use of the apsidal motion constant which amounts to half the tidal Love number used to estimate the tidal quality factor \qstar \citep[cf.][]{MardlingLin02}.}
\begin{equation}\label{eq:tstar}
\tstar = {9\over 4} {\mpl^2 \over \left(\mpl+\mstar\right)} \wpl \left(\wpl - \wstar\right) {\rstar^5\over {\qstar a^3}}  
\end{equation}
where 
the migration rate of the planet due to the tidal interaction with the star is given by:
\begin{equation}\label{dadtpl}
\left({\textrm{d}a\over \textrm{d}t}\right)_\star = \textrm{sign}(\wstar - \wpl) {9\over 2} \left({G \over {a\mstar}}\right)^{1/2} \left({\rstar\over a}\right)^5 {\mpl \over \qstar}
\end{equation}
where $ \wpl$ is the planet orbital rotation rate, \qstar\ is the tidal dissipation factor \cite[e.g.][]{ZhangPenev14}. The sign of the migration depends on whether the planet orbits within or beyond the corotation radius. The migration induced by the disk is always inwards
\begin{equation}\label{dadtdisk}
\left({\textrm{d}a\over \textrm{d}t}\right)_{disk} = -{a\over\taumig} .
\end{equation}
 In order to slow the star down, the planet must orbit beyond the corotation radius. Hence, the planet's inward migration must be stopped before it reaches the corotation radius. We must then have
\begin{equation}
\left({\textrm{d}a\over \textrm{d}t}\right)_\star \geq \left|\left({\textrm{d}a\over \textrm{d}t}\right)_{disk}\right|
\end{equation}
for $r \geq \rco$ all the way down to the corotation radius. Combining equations~\ref{dadtpl} and \ref{dadtdisk}, we compute the critical semi-major axis at which the disk and stellar torques have the same amplitude on the planet \citep[see also][]{Lin96},
\begin{equation}\label{acrit}
\acrit = \left({{9\mpl\taumig}\over{2\qstar}}\right)^{2/13}\left({G\over\mstar}\right)^{1/13}\rstar^{10/13} .
\end{equation}
At a semi-major axis larger than \acrit\ the planet will migrate inwards, while for distance smaller than \acrit\  the migration will be outwards.  In order to ensure outwards migration at or beyond the corotation radius, as required to brake the central star, \acrit\ must thus to be larger than \rco. Using parameters listed in Table~\ref{param}, and \qstar = 10$^6$, one finds \acrit = 2.4 10$^{11}$ cm = 1.7\rstar = 0.3\rco. The magnitude of the tidal torque  thus appears unable to prevent the planet from migrating inward the corotation radius, at which point it will contribute to spin up the star. Indeed, the tidal torque decreases very steeply with distance as seen from Eq.~\ref{dadtpl}.  Its amplitude at the corotation radius is about 4 orders of magnitude weaker than required to balance the migration and accretion torques (cf. Fig.~\ref{torq}), a result which is barely sensitive to the parameters of the system (cf. Eq.~\ref{acrit}).

\subsection{Magnetic torques}

\subsubsection{Models and kind of magnetic interactions}
In this section, we aim at estimating orders of magnitude of the magnetic torques exerted by the planet on the star. To do so, we model the planet and the star as objects with a given radial profile of electrical conductivities, possibly generating magnetic fields. Naturally, this also models magnetic interactions between a moon and its planet, such as Io and Jupiter. The following is thus valid for star-planet or moon-planet magnetic interactions.

In the simplest case considered by some authors \cite[][]{Laine2008,Chang2010,Chang2012}, vacuum is assumed between the star and the planet. If the star and the planet both generate a dipolar field, the magnetic torque is simply given by the cross product of the magnetic moment with the magnetic field produced by the other dipole (which tends to align the magnetic moments). Torque can also come from dissipative effects. Typically, any time-variation of the magnetic field in an electrically conductive domain will generate eddy currents, leading to Joule dissipation and torques. This is the so-called Transverse Electric (TE) mode \cite[][]{Laine2008}, and the associated torques are e.g. studied by \cite{Chang2012} in the case of a planet orbiting in a tilted stellar magnetic field, which generates eddy currents in the conducting planet.

Actually, the space between the star and the planet is rather filled by a stellar wind originating from the star, which can be roughly modeled as an electrically conducting fluid in motion. This has, in particular, two consequences: (i) the stellar magnetic field is advected by the stellar wind, which is the so-called interplanetary magnetic field (IMF), and decays thus less rapidly than in vacuum, and (ii) waves and currents can exist between the star and the planet allowing new kinds of magnetic interactions. The former point leads to a stronger TE mode in the planet, due to the higher time-varying magnetic field advected by the stellar wind. The latter point requires to investigate the interaction between the planet and the surrounding magnetized flow of the stellar wind.

The star-planet interaction via the stellar wind can be of various kinds, depending of the planet's velocity $\mathbf{v_{orb}}$ relative to the stellar wind (SW) local one $\mathbf{v_{sw}}$, the speed $v_{A,f}$ of the fastest wave in the stellar wind (i.e. the so-called fast magnetosonic wave), and the speed $v_{A}$ of the shear (or intermediate) Alfv\'en wave. First, if the (fast) Alfv\'en Mach number $M_{A,f}=||\mathbf{v_{orb}}-\mathbf{v_{sw}}||/v_{A,f}$ is larger than $1$, the obstacle constituted by the planet generates a shock wave. In this so-called super-Alfvenic case, the flow is controlled upstream and disturbances are transmitted downstream. This is the case for Venus for instance, where the shock wave takes the usual form of a (detached) bow shock due to its bluff (i.e. non sharp-nosed) geometry.  In the case where the planet generates its magnetic field, like the Earth, this changes the apparent radius of the obstacle (aka the magnetosphere radius) for the stellar wind, leading to a enhanced coupling. This case of magnetic interaction between the stellar wind and the planet is called magnetospheric interaction by \cite{Zarka2007a}.

Second, we consider the sub-Alfvenic case ($M_{A,f}<1$), focusing first on the simple case where the planet is weakly magnetized or unmagnetized. In this case, the planet motion generates shear (or intermediate) Alfv\'en waves which propagate in the stellar wind and transport some energy from the planet to the star. The wave can then be reflected back, from the star to the planet, which leads to a planet-star interaction. Since the disturbance is only partially reflected, each subsequent reflection is of lower amplitude, and results in a lesser change to the current system \cite[][]{Crary1997}. After several round-trip travel times, the system reaches a steady state, and a current loop is settled. In this case, the star-planet interaction can be modeled as a DC circuit \cite[][]{Goldreich1969}, which is called the unipolar inductor model or the TM mode \cite[][]{Laine2013}. In the other limiting case, the planet has moved away when the wave comes back and the planet's interaction is thus decoupled from the star. This is the so-called Alfv\'en wings (or Alfvenic interactions)  model \cite[e.g.][]{Saur2004a}. These two models, sometimes presented as two different kinds of interaction, originates actually from the same phenomena and have been primarily applied to the Io-Jupiter interaction, but also to planets around magnetic dwarfs, ultra compact WD binaries, exoplanetary systems, etc.

In the regime $M_{A,f}<1$, we finally have the case where both the planet and the star generate a magnetic field. In the stellar system, the only example of such a case is the Ganymede-Jupiter interaction, which constitutes a textbook example of the expected plasma environment around close-in extrastellar planets \cite[][]{Saur14}. This kind of interaction is often termed the dipolar interaction \cite[e.g.][]{Zarka2007a}.

\subsubsection{A generic torque formula}
Focusing on dissipative torques, one can link the dissipated power $P_d$ with the star-planet dissipative torque $T$ by 
\begin{equation}
P_d=T ||\mathbf{v_{pl}}-\mathbf{v^{B}}||/a, \label{eq:Pd}
\end{equation}
where $a$ is the star-planet distance, $\mathbf{v_{pl}}$ is the planet velocity ($v_{pl} \approx  \wpl a$) and $\mathbf{v^{B}}$ is the advective velocity of the magnetic field at the planet orbit (i.e. the SW velocity $\mathbf{v^{B}}=\mathbf{v^{sw}}$, or $v^{B}=\wstar a $ in absence of SW flow). Dimensional arguments allow to write any torque exerted on the planet by the star as
\begin{equation}
T=\epsilon\, a\, A_{eff}\, P_{eff}^{ram},
\end{equation}
where $\epsilon$ is a dimensionless coefficient, $a$ is the star-planet distance, $A_{eff}$ is the typical surface area on which the star-planet magnetic coupling is effective, and $P_{eff}^{ram}$ is the typical energy density of the coupling on the area $A_{eff}$  (which can be seen as the ram pressure of the stellar wind when a stellar wind is considered). Focusing only on the magnetic coupling, this formula reduces to 
\begin{equation}
T=\epsilon\, a\, A_{eff}\,  \frac{B_{eff}^2}{\mu_0} =\epsilon\, a\,  \pi R_{obs}^2\, \frac{B_{eff}^2}{\mu_0} , \label{eq:Torq}
\end{equation}
where $B_{eff}$ is a typical magnetic field value exerted on the area $A_{eff}$, $\mu_0$ is the magnetic permeability of the vacuum, and $R_{obs}=\sqrt{A_{eff}/\pi}$ is the typical radius of $A_{eff}$.

Equation (\ref{eq:Torq}) requires the knowledge of $\epsilon$,  $B_{eff}$ and $A_{eff}$. The expression of $\epsilon$ depends on the magnetic interaction we consider, which is discussed in section \ref{sec:formula}. The calculation of magnetic field $B_{eff}$ depends on the medium between the planet and the star. If this is vacuum, then $B_{eff}$ is simply deduced from the Gauss coefficients of the stellar magnetic field. If a stellar wind exists, $B_{eff}$ is simply given by the model chosen for the stellar wind. Finally, $A_{eff}$ depends on the planet. In absence of any external conductive layer, $R_{obs}=\rpl$, which is thus a lower bound for $R_{obs}$. However, a planet has usually an ionosphere due to photoionization of the upper atmosphere (even without any own magnetic field), which leads to an induced magnetosphere above the ionosphere. This also means that the time-varying stellar magnetic field can be shielded by the ionosphere such that induced eddy currents in the planetary interior is small. Note that, as argued by \cite{Chang2012},  atmosphere circulation may maintain this so-called exo-ionosphere on the permanent nightside of a tidally locked planet. In this case, the radius $R_{obs}$ to consider is the exo-ionospheric radius, which is typically $R_{iono}=1.1-1.4\, \rpl$ in the case of Io \cite[][]{Zarka2007a}. Finally, the planet can also have a own magnetic field, either stored in its rocks, or generated by dynamo. In this case, the planet develops its own magnetosphere. The magnetopause radius $R_{MP}$ is then the apparent radius $R_{obs}$ to consider, which can be estimated by the balance between the total ram pressure $p_{ram}$ coming from the star and the planetary magnetic field $\bpl$:
\begin{equation}
R_{MP} = \rpl \left[ \frac{\bpl^2}{2\, \mu_0\, p_{ram}} \right]^{\frac{1}{2m}}, \label{eq:RobsM0}
\end{equation}
where $m$ depends on the planetary magnetic field geometry ($m=3$ for a mainly dipolar magnetic field, $m=4$ for a mainly quadrupolar magnetic, field, etc.), and where
\begin{equation}
p_{ram}=\frac{1}{2} \rho\, ||\mathbf{v_{pl}}-\mathbf{v^{sw}}||^2+\frac{1}{2} \frac{B_{eff}^2}{\mu_0}.
\end{equation}
Finally, $R_{obs}$ is thus simply given by
\begin{equation}
R_{obs}=\max (\rpl,R_{iono},R_{MP}). \label{eq:RobsM}
\end{equation}
Particular forms of the generic formula (\ref{eq:Torq}) has already been considered in previous works. For instance, neglecting $\mathbf{v_{pl}}$ in equation (\ref{eq:Pd}) allows to recover equation 9 of \cite{Zarka2007a}, where $B_{eff}=B_{\perp}$ is the IMF component perpendicular to the SW flow in the planet's frame. Similarly, equation (9) of \cite{Chang2010} is recovered with equation (\ref{eq:Torq}), assuming no stellar wind and a dipolar stellar field, such that $B_{eff}=\bstar(\rstar/a)^3$, and considering $R_{obs}=\sqrt{2}\, \rpl$. 


As stated by \cite{Zarka2007a}, this general expression is simply the fraction $\epsilon$ of the magnetic energy flux convected on the obstacle (aka the planet), and is expected to provide a correct order of magnitude whatever the interaction regime (unipolar or dipolar, super- or sub- Alfvenic) as long as the obstacle conductivity is not vanishingly small. The physics of the interaction is now hidden in the coefficient $\epsilon$, which is the difficult estimate we have to obtain.

\subsubsection{Application of the formula} \label{sec:formula}

\begin{figure*}
   \centering
   \includegraphics[width=18cm]{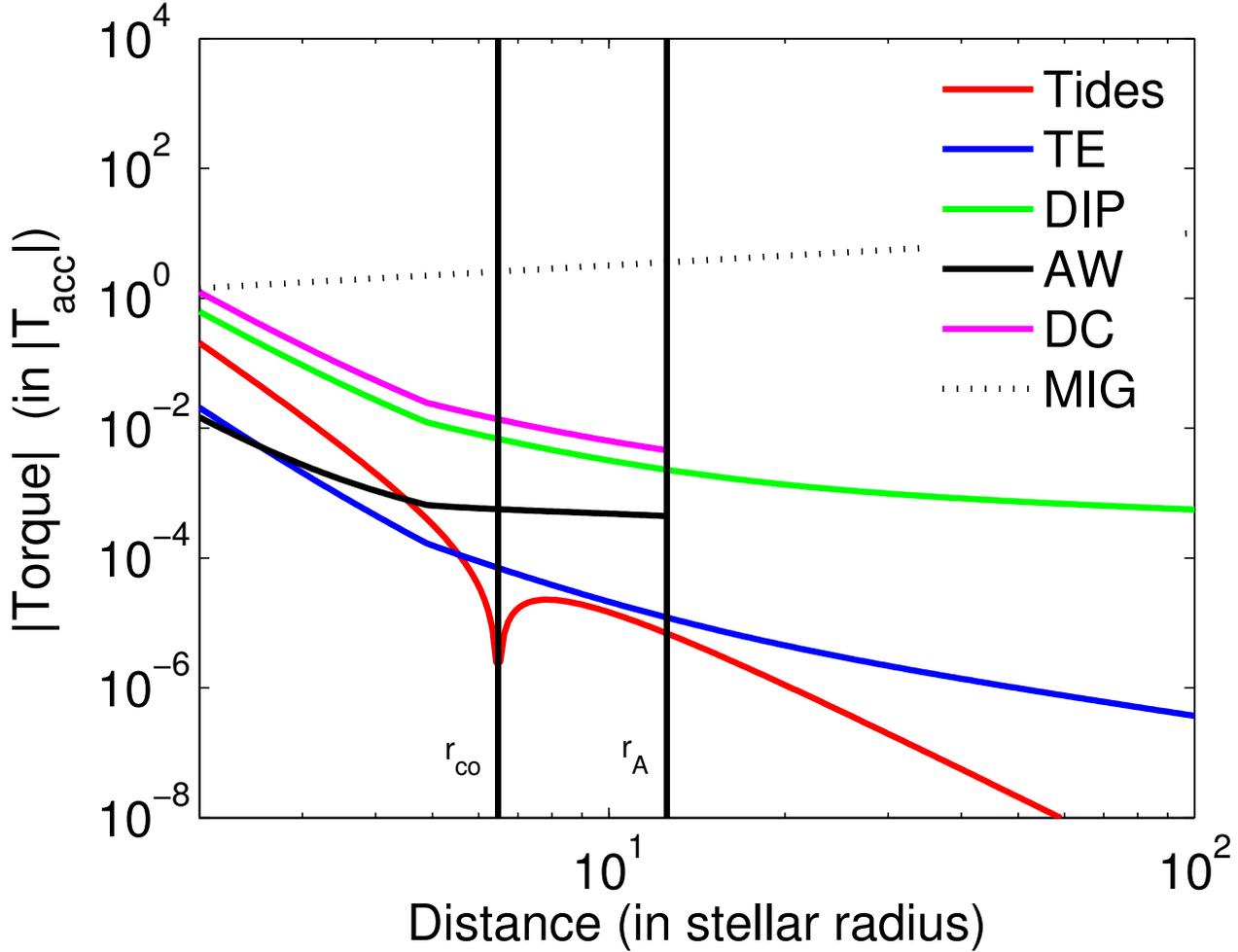}
   \caption{Magnitude of the torques (in accretion torque units). Considering the left vertical axis, the curves show the magnetic torque of the TE mode (eq. \ref{eq:TEm} and $R_{obs}=\rpl$), of the Alfven wings (eq. \ref{eq:Cdm} and $R_{obs}=\rpl$), the dipolar interaction ($\epsilon=1$, eq. \ref{eq:RobsM} for a  dipolar planetary surface magnetic field of $10\,\textrm{G}$), the tidal torque, the magnetic torque  associated with the unipolar inductor model (eq. \ref{eq:Laine}, with a planetary conductivity of  $\sigma_{pl}=10^{-2} $ S/m.), and the migration torque ($\tau_{mig}=10^6\, \textrm{yr}$), shown by a dotted line. From left to right, the solid lines are respectively the corotation and the Alfven radii, whereas the vertical dashed line shows $r_v$, the upper radius of the favorable DC circuit closure regime. We have also plotted, for comparison, the empirical formula obtained by Strugarek et al. (2014) from 2.5D numerical simulations. Note also the change of slope noticeable at a distance of 5 \rstar\  in some curves, which is due to the stellar wind becoming unable to prevent the magnetic planet to develop an extended magnetosphere. Only the magnitude of the torques is plotted, not their sign, which is the reason why the tidal torque (cf. Eq.~\ref{eq:tstar}) does not change sign at the corotation radius in this plot.} 
         \label{torq}
\end{figure*}

In this work, we aim at studying if the star can be slowed down by planetary magnetic torque. Since the super-Alfvenic regime only allows disturbances to be transmitted downstream, this requires to consider sub-Alfvenic regime such that a torque can be exerted by the planet on the star. Given that the stellar wind accelerates when flowing away from the star, this imposes to consider orbital distance $a$ smaller than the so-called Alfv\'en radius given by \cite[][]{Bessolaz2008}
\begin{equation}\label{eq:ra}
\ra = \left({{\bstar^4\rstar^{12}}\over{2G\mstar\macc^2}}\right)^{1/7} = 12\rstar,
\end{equation}
distance at which the SW flow becomes super-Alfvenic\footnote{ Using the expression of the Alfv\'en radius from \citet{MattPudritz08}'s stellar wind models (their Eq.14) would result in twice as large an estimate.}. We have thus to estimate torques in four cases, the TE mode, the Alfven wings model), the dipolar interaction, and the unipolar inductor model. 
  
\bigskip
\noindent{\bf Torque due to the TE mode: } \label{sec:TE}
In this case, the torque is associated with the Joule dissipation in the interior of the planet, originating from the stellar magnetic field oscillating at the rate $\omega$. Owing to the diffusive nature of the problem, one can thus estimate the Ohmic dissipation by \cite[][]{Campbell,Campbell1983,Laine2008,Chang2012}:
\begin{equation}
P_{ohmic}=\mathcal{V}\, \omega\ \frac{B_{eff}^2}{2 \mu_0},
\end{equation}
where $\mathcal{V}$ is the volume where eddy currents take place. This is thus typically a thin shell of thickness $\delta$, which gives $\mathcal{V}=A_{eff}\, \delta$, i.e. $ \mathcal{V} \approx 4 \pi \rpl^2 \delta$ if it takes place at the surface of an electrically conductive planet. Noting $\eta$ the typical planetary electrical diffusivity, $\delta=(2 \eta/\omega)$ is then the skin depth for our magnetic induction problem.

Finally, considering that  $\omega \sim \wpl$, an upper bound of the torque is obtained for $\mathcal{V}=4 \pi /3 \cdot \rpl^3$, and we will thus use equation (\ref{eq:Torq}) using $R_{obs}=\rpl$ and
\begin{equation}
\epsilon=\frac{2}{3} \frac{\rpl}{a}. \label{eq:TEm}
\end{equation} 

\bigskip
\noindent{\bf Alfven wings:} \label{sec:Alfven}
As shown by \cite{Drell1965}, the radiation of two pure linear Alfven wings leads (when converted in S.I. units) to a torque given by formula (\ref{eq:Torq}), with the coefficient \cite[see also][]{Lai2012}
\begin{equation}
\epsilon=2\, M_A, \label{eq:Drell}
\end{equation}
where $M_{A}=||\mathbf{v_{orb}}-\mathbf{v_{sw}}||/v_{A}$ is the shear (or intermediate) Alfv\'en wave Mach number, with $v_A=B^2/\sqrt{\rho \mu_0}$ is the shear (or intermediate) Alfv\'en wave velocity and $\rho$ the stellar wind density. The factor $2$ in equation (\ref{eq:Drell}) naturally comes from the existence of two Alfven wings. The work of  \cite{Drell1965} has then been extended by \cite{Neubauer1980} to the fully non-linear sub-Alfvenic situation, also including flow $\mathbf{v_{orb}}-\mathbf{v_{sw}}$ which is not perpendicular to the background field  $\mathbf{B}$. Following \cite{Zarka2007a}, and taking into account the existence of $2$ wing, this leads to the extended formula 
\begin{equation}
\epsilon=\frac{2\, M_A}{||M_A\, \mathbf{\hat{v}_{rel}}+ \mathbf{\hat{B}} ||}, \label{eq:Cdm0}
\end{equation}
with the two unit vectors $ \mathbf{\hat{v}_{rel}}$ and $ \mathbf{\hat{B}} $, directing respectively the relative velocity $\mathbf{v_{orb}}-\mathbf{v_{sw}}$ and the (background) magnetic field $\mathbf{B}$. Note that we naturally recover equation (\ref{eq:Drell}) for $M_A \ll 1$. Focusing on orders of magnitude, the factor $2$ can be dropped, and the case where the flow is perpendicular to the magnetic field ($\mathbf{\hat{v}_{rel}} \cdot \mathbf{\hat{B}} =0$) leads then to \cite[][]{Zarka2007a}
\begin{equation}
\epsilon=\frac{1}{\sqrt{1+1/{M_A}^{2}}}, \label{eq:Cdm}
\end{equation}
for the Alfven wings interaction. Note that, in this case, $R_{obs}$ is of the order $R_{obs} \sim \rpl$,  even in the presence of an exo-ionosphere. Note also that Alfven waves cannot be radiated when $r>r_A$, and the Alfven wings interaction is thus only valid for $r<r_A$.

\bigskip
\noindent{\bf Dipolar or magnetospheric interaction:} \label{sec:dipolar}
As argued by \cite{Zarka2007a}, the torques associated to these interactions can also be modeled by formula (\ref{eq:Torq}), using $\epsilon=0.1-1$ (typically, $\epsilon\approx0.3$ in the sub-Alfvenic Ganymede-Jupiter interaction). Note that $\epsilon=1$ is in agreement with equation (\ref{eq:Cdm}), which gives $\epsilon \sim 1$ for $M_A \geq 1$. The important feature of this kind of interaction is the presence of a own planetary magnetic field, which can lead to large magnetic interactions (due to large $R_{obs}$).

\bigskip
\noindent{\bf Unipolar inductor model (DC circuit or TM mode):} \label{sec:uni}
In this case, first studied by \cite{Goldreich1969}, the Alfv\'en waves round trips between the star and the planet allow to reach a steady state, which can be modeled as a DC circuit. In the planetary frame, the electrical field $\mathbf{E}=\mathbf{v} \times \mathbf{B}$ generated by the planetary velocity $\mathbf{v}$ allows indeed to close a DC circuit through the surrounding electrically conductive medium. An electrical current is thus flowing through the electrical resistances of the planet, $\mathcal{R}_p$, of the stellar footprint of the planet, $\restar$, and of the two flux tubes crossing the stellar wind, $2 \mathcal{R}_f$. 

The difficult estimation of the three electrical resistances is performed here by following \cite{Laine2012,Laine2013}, who neglect $\mathcal{R}_f \ll \mathcal{R}_p,\, \restar$. According to the formula obtained by \cite{Laine2012,Laine2013}, we thus end up with
\begin{equation}
\epsilon=\frac{8\  \mu_0\, a\, |\wpl-\wstar| }{\pi\, (\mathcal{R}_p+\restar)} \label{eq:Laine}
\end{equation}
One can distinguish different regimes of closures of the DC circuit \cite[][]{Laine2013}. In the so-called favorable DC circuit closure regime, the Alfven wave have the time to perform the round-trip planet-star-planet, which actually imposes here that the planet is on an orbit below a certain distance $r_v$.  In our model, $r_v \simeq 8 \rstar$. Analytical estimates of $r_v$ are derived in Appendix A.

Note that the DC circuit is naturally opened for $r>r_A$. Note also that it exists an upper limit to the magnetic interaction torque generated using the unipolar inductor model \cite[][]{Lai2012}. Indeed, when the circuit resistance is too small, the large current flow severely twists the magnetic flux tube connecting the two binary components, leading to the breakdown of the circuit. According to \cite{Lai2012}, this breakdown appears when 
\begin{equation}
\zeta=\frac{4\, \mu_0}{\pi} \frac{ a|\wpl-\wstar|}{ \mathcal{R}_p+\restar}=\frac{\epsilon}{2}  \geq 1,
\end{equation}
which limits $\epsilon$ to $\epsilon=2$ \cite[corresponding to $\zeta=1$ in][]{Lai2012}.

\subsubsection{Results} 

In order to calculate the various torques estimates give above, a
stellar wind model is required. In this work, we have considered two
stellar wind models. One is adapted from \cite{Zarka2007a}, as
detailed in Appendix~B, and the other one is described in Lovelace et
al. (2008). Both give similar results as far as magnetic torques are concerned (cf. Appendix~B). Using the stellar wind model adpated from  \cite{Zarka2007a}, the results are summarized in Figure \ref{torq}, and compared with the tidal torque. Note that the magnetic torque associated with the DC circuit depends on the chosen planetary conductivity. It turns out that above $\sigma_{pl}=10^{-5} $ S/m (as used in Fig. \ref{torq}), the resistance is very small ($\zeta \geq 1$), and the torque is thus maximum, given by $\epsilon=2$. 


Fig.~\ref{torq} clearly illustrates the main result of this study. None of the planet-induced torques acting on the central star, be it tidal or magnetic, is able to balance the accretion torque beyond the corotation radius. Hence, at least with the parameters adopted here, the star-planet interaction cannot prevent the spin up of the central star as it accretes from its circumstellar disk and contracts down its Hayashi track. Indeed, even the most powerful torques, corresponding to the dipolar interaction and the unipolar inductor model, fail by several orders of magnitude to match the magnitude of the migration torque beyond the corotation radius.  We have also compared these results to \citet{Strugarek14} 2.5D numerical simulations and find that their parametric torque formulation predicts order of magnitude torques similar as the DC and dipolar cases investigated here (see Appendix~\ref{sec:appendixC} for a short account of their parametric torque formulation).  

One may wonder why the magnetic coupling seems to be efficient enough in the similar study of \citet{Fleck08}, in contrast to our results. This is partly due to the fact that he prescribes the cross-sectional area $A_{eff}$ of the magnetic flux tube linking a planet to its host star to be in the range $\alpha=1-10\%$ of the stellar surface. Here, we determine $A_{eff}$ self-consistently and show that it is of order of magnitude of $\pi R_{obs}^2$, where $R_{obs}\approx\rpl$ when the planet is close from the star, and thus $\alpha \approx \pi\rpl^2/4\pi\rstar^2 \approx 0.2 \%$ with our values. Also, \citet{Fleck08} assumed a stronger magnetic field (2 kG) and a larger stellar radius (3\rsun) than we adopted here, which results in a much stronger torque (cf. his Eq. 14). Indeed, a 3 kG field associated to a 4 \rsun\  protostar would be required to obtain a torque comparable to the accretion and migration torques. 

\section{Discussion}

We have considered the various torques acting in a system consisting
of a contracting pre-main sequence star still accreting from its
circumstellar disk into which a newly-formed Jupiter-mass planet is
embedded on a short-period orbit. We have shown that a balance between
accelerating torques acting onto the star, due to accretion and
contraction, and decelerating ones, due to the star-planet interaction,
cannot be reached under the conditions we explored here. For all cases we investigated, the accretion torque exceeds any planetary torque acting on the central star by orders of magnitude. 

The main limitation of the scenario outlined here lies in the adopted
wind model. The model is an extrapolation from a solar-type wind
properties, which is not necessarily adapted to the wind topology of
young stars. Firstly, the topology of the magnetic field of young
stars might be quite different from that of a mature solar-type stars
\cite[e.g.][]{DonatiLandstreet09} , although current measurements
suggest that fully convective PMS stars at the start of their Hayashi
tracks host strong dipolar fields, of order of a few kG \cite[][]{Gregory12, Johnstone14}. Secondly, as shown by \citet[][]{ZanniFerreira09}, the structure of the stellar magnetosphere
might be significantly impacted by its interaction with the
circumstellar disk. In particular, an initially dipolar magnetic field
may evolve into a much more complex and dynamical topology. Whether
this would significantly modify the torques associated to the magnetic
star-planet interaction is yet unclear. 

Another issue of course is whether the framework proposed here is applicable to the
vast majority of young stars that rotate slowly as they appear in the
HR diagram. This would require that planet formation be not only a
common occurrence, but also that it is fast enough to impact the
earliest stages of stellar evolution, and that it is quite dynamic so
as to frequently send massive planets on inner orbits. Whether all
these conditions are met around protostars is yet unknown. The recent
ALMA image of the numerous gaps in the disk of the protostar HL Tau
suggests that multiple planet formation may indeed proceed quite
rapidly after protostellar collapse. The significant fraction of hot
Jupiters found to revolve around their host star on a non coplanar
orbit further suggests that dynamical interaction between forming
planets in the protostellar disk may be efficient to scatter massive
planets close to their parent star \cite[][]{FabryckyTremaine07, Triaud10, MortonJohnson11,
Albrecht12, CridaBatygin14}. Eventually, Hot Jupiters (HJs)
are deemed to coalesce with the central star after a few hundred
million years \cite[][]{Levrard09, Bolmont12}. Hence, the
scarcity of HJs around mature solar-type stars does not
necessarily exclude that they are much more common during early
stellar evolution \cite[e.g.,][]{TeitlerKonigl14, Mulders15}. 


Finally, we wish to emphasize that the preliminary results presented
here have to be extended in order to explore a much larger parameter
space. 
Table~\ref{param} summarizes typical parameters
for young accreting stars that we adopted here for the torque
computations. However, classical T Tauri stars at a given mass and age
exhibit a wide range of spin rates \cite[][]{Bouvier14}, mass
accretion rates \cite[][]{Venuti14}, magnetic field strength and
topology \cite[][]{Donati13}.  Hence, the torque estimates provided here may
vary significantly from one system to the next. Moreover, during the
embedded phase, other sets of parameters could be relevant, implying
for instance a larger radius, and possibly stronger magnetic fields \cite[e.g.,][]{Grosso97, Tsuboi00}
 and faster rotation
rates \cite[e.g.,][]{Covey05}. Whether the planetary lift scenario could be
instrumental in the protostellar embedded phase to produce the low
angular momentum content of revealed young stars remains to be
ascertain. Also, more complex scenarios than that described here could be envisioned. For instance, the planetary migration torque could be significantly reduced, and the migration timescale considerably lengthened, by tidal resonances occurring between several inner planets embedded in the disk, or by considering a steeper surface density profile in the disk close to the magnetospheric cavity. In the latter case, the enhanced outward flux of angular momentum between the inner disk material and the close-in planet would result in sub-Keplerian disk rotation close to the truncation radius, thus potentially reducing the accretion torque onto the central star.


\section{Conclusion}

As an alternative, or a complement, to star-dik interaction models that attempt to account for the low spin rates of newly-born stars,
 we explored here the tidal and magnetic interactions between a
young magnetic star and a proto-hot Jupiter embedded in the inner disk. We investigated whether
such a close-in embedded planet could extract enough angular momentum from the central star to
counteract both the accretion and contraction torques, thus leaving
the star at constant angular velocity as it
evolves on its Hayashi track. While a lot remains to be done in exploring
the parameter space relevant to such systems, it appears that the decelerating torques acting on the central star are orders of magnitude too weak to counterbalance the spin up torques. Hence, even though planetary systems may promptly form around embedded young stars, as suggested by the recent ALMA image of HL Tau's disk\footnote{cf. http://www.eso.org/public/unitedkingdom/news/eso1436/}, star-planet interaction may not eventually prove a viable alternative to magnetic star-disk interaction to understand the origin of the low angular momentum content of young stellar objects.


\appendix 

\section{Analytical estimates of $r_v$ for the unipolar inductor model}

Based on the work of \cite{Laine2013}, we have been able to obtain analytical estimates of $r_v$, the transitional radius between favorable and unfavorable DC circuit closure regimes (see. Eq.~\ref{eq:Laine}). Indeed, noting $t_0=2\, \rpl /||\mathbf{v_{pl}}-\mathbf{v^{sw}}||$ the advection time  of the unperturbed plasma to cross the planet's diameter, and $t_{A,FT}$ the Alfven wave travel time between the planet and the top of the footprint, the condition for favorable closure regime reads then $t_0\geq 2 t_{A,FT}$. Estimating $ t_{A,FT}$ by the time  $t_{A,FT0}$ it takes for the Alfven wave to travel along the flux tube until the stellar chromosphere, we obtain the following asymptotic estimates:
\begin{equation}
r_v=\rstar \left(\frac{3 \bstar\, \rpl}{\rstar \wstar} \sqrt{\frac{4 \pi}{\mu_0 } \frac{ V_{*}}{\mwind}} \right)^{1/4}, \label{eq:rv1}
\end{equation}
 for slow rotators ($\wpl \ll \wstar$ at $r=r_v$), and 
\begin{equation}
r_v=\rstar \left(\frac{9 \bstar^2\,  \rpl^2\, \rstar }{G \mstar} \frac{4 \pi}{\mu_0 } \frac{ V_{*}}{\mwind} \right)^{1/5}, \label{eq:rv2}
\end{equation}
for fast rotators ($\wpl \gg \wstar$ at $r=r_v$).

\section{Stellar wind model adapted from Zarka (2007)}

The star we consider being not very different from the Sun, we have chosen to slightly adapt the solar wind model proposed by \cite{Zarka2007a}. First, let remind roughly the various usual scaling laws at play in the solar wind, before detailing the exact model used in this work. At a certain distance $r$ to the Sun (beyond a few Sun radius), the radial SW velocity can be considered as constant, and thus, mass and magnetic flux conservation give a SW density $N$ and a radial field $B_r$ decaying as $1/r^2$,  whereas other conserved quantities (e.g. angular momentum) give an azimuthal velocity $v_{\phi}^{sw}$ and magnetic field $B_{\phi}$ which decays as $1/r$  \cite[e.g.][]{Weber1967,Belcher1976}. Close from the Sun, $B_r$ is rather dipolar and decays thus as $1/r^3$ (which leads to a decay in $1/r^2$ of $B_{\phi}$, see eq. \ref{eq:A2}).

The model proposed by \cite{Zarka2007a} for the Sun and the solar wind is (in spherical coordinates)
\begin{eqnarray}
B_r&=&\frac{\lambda}{(r/\rstar)^2} \left(1+\frac{f-1}{(r/\rstar)^{3/2}} \right), \\
B_{\phi}&=&B_r\, \frac{\Omega\, r}{||\mathbf{v_{pl}}-\mathbf{v^{sw}}||}, \label{eq:A2} \\ 
B_{\theta}&=&0
\end{eqnarray}
with the constants $4 \leq f \leq 9$ (here taken equal to $f=6.5$) and $\lambda \approx 1.5 \times 10^{-4}\, \textrm{T}$, and
\begin{eqnarray}
N_e(\textrm{cm}^{-3})=\alpha_1\, (a/\rstar)^{-15}+\alpha_2\, (a/\rstar)^{-9/2}+\alpha_3\, (a/\rstar)^{-2} 
\end{eqnarray}
with $\alpha_1 \approx 3\times 10^8\, \textrm{cm}^{-3}$,  $\alpha_2 \approx 4\times 10^6 \, \textrm{cm}^{-3}$, and  $\alpha_3 \approx 2.3\times 10^5\, \textrm{cm}^{-3}$, which gives the SW density $\rho=1.1\, m_p\, N_E$ (with the proton mass $m_p\approx1.673 \times 10^{-27}\, \textrm{kg}$). In this model, the solar wind velocity $\mathbf{v_{sw}}$ can be described by
\begin{eqnarray}
v_r^{sw}&=& v_{\infty}^{sw} \left[ 1-\exp \left(- \frac{r/\rstar-1.45}{5} \right) \right] \\
v_{\phi}^{sw}&=&0, \\
v_{\theta}^{sw}&=&0,
\end{eqnarray}
with $v_{\infty}^{sw}=385\, \textrm{km/s}$. Note that the values given by this expression of $v_r^{sw}$ are naturally quite close from the ones derived from $\rho$ by mass conservation (i.e. using $r^2\, \rho\, v_r^{sw}=cst$).

We have adapted this model to our star by multiplying $N_e$ by $\mwind/\rstar^2 \sqrt{\rstar/\mstar}$, the magnetic field by $ \bstar$, and $\mathbf{v_{sw}}$ by  $\sqrt{\mstar/\rstar}$, all these quantities being expressed in solar units \cite[which allows to recover the initial solar wind model of][]{Zarka2007a}, i.e. $\rsun=6.96 \times 10^8\, \textrm{m}$, $\msun=1.99 \times 10^{30}\, \textrm{kg}$, $2 \pi/\wsun=27\, \textrm{d}$, $\mwsun = 1.6 \times 10^{-14}\, \msunyr$, and $\bsun=10\,  \textrm{G}$ \cite[these two latter values are actually obtained as outputs from the solar wind model of][]{Zarka2007a}.

Note finally that theoretical models, such as the one of \cite{Weber1967}, rather advocates an azimuthal magnetic field given by $B_{\phi}=B_r(v_{\phi}^{sw}-\wstar r)/v_r^{sw}$, as well as non-zero azimuthal velocity, typically given by co-rotation ($v_{\phi}^{sw}=\wstar r$) for $r<r_A$ and $v_{\phi}^{sw}=\wstar r_A (r_A/r)$ for $r>r_A$. We have checked that this does not change our conclusions, e.g., when using the stellar wind model of \cite{Lovelace08}. This is expected since, in our case, $v_{\phi}^{sw}$ is at most of the order of $ v_r^{sw}$, or smaller, which would thus even further reduce the magnetic torques.

   \begin{figure}
   \centering
   \includegraphics[width=9cm]{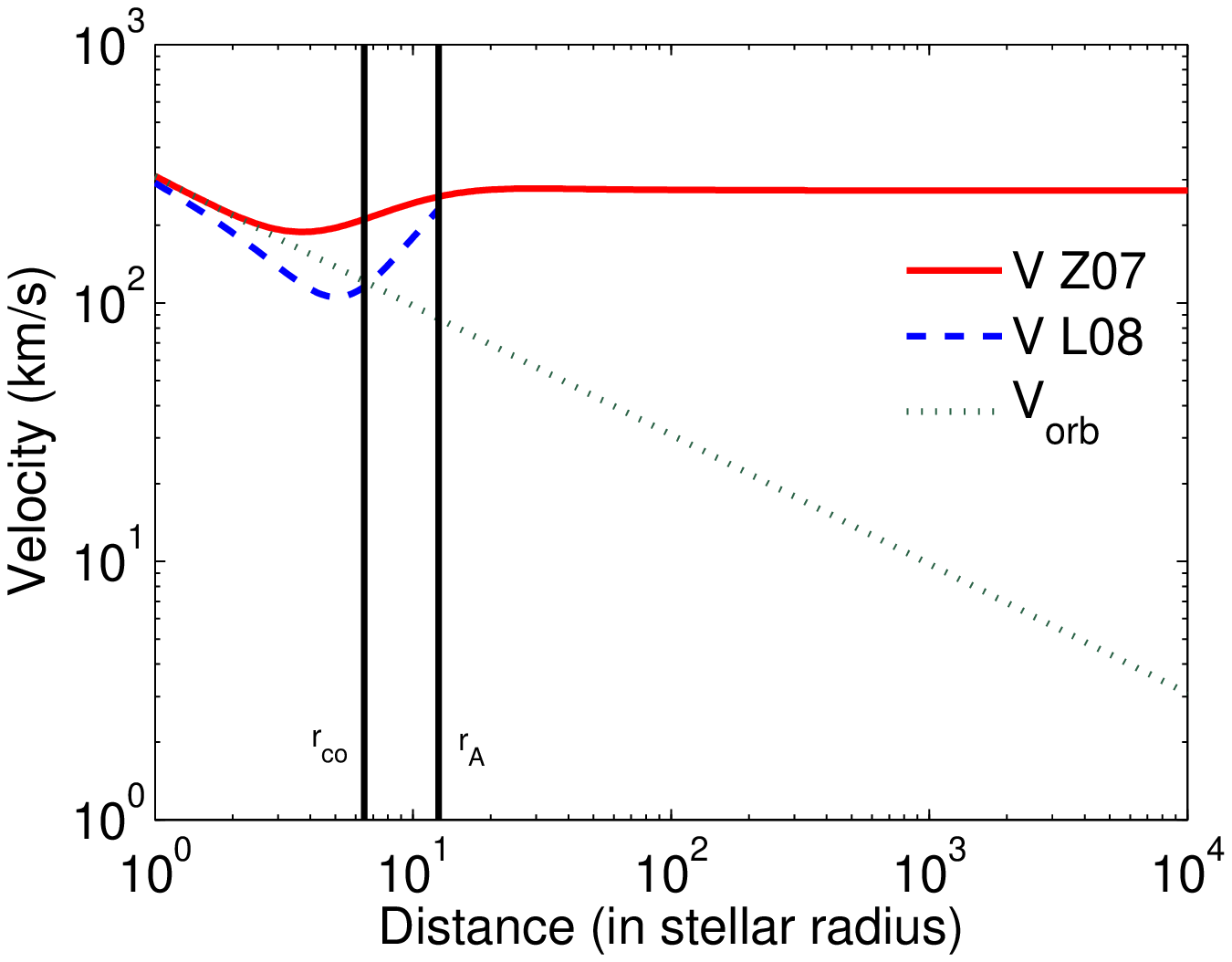}
      \caption{Comparison of the stellar wind speed $V=||\mathbf{v_{sw}}-\mathbf{v_{orb}}||$ in the planetary frame for the model adapted from Zarka (2007), labeled Z07, and Lovelace et al. (2008) labeled L08. The planetary orbital speed $V_{orb}$ has also been plotted. }
         \label{speed}
   \end{figure}

   \begin{figure}
   \centering
   \includegraphics[width=9cm]{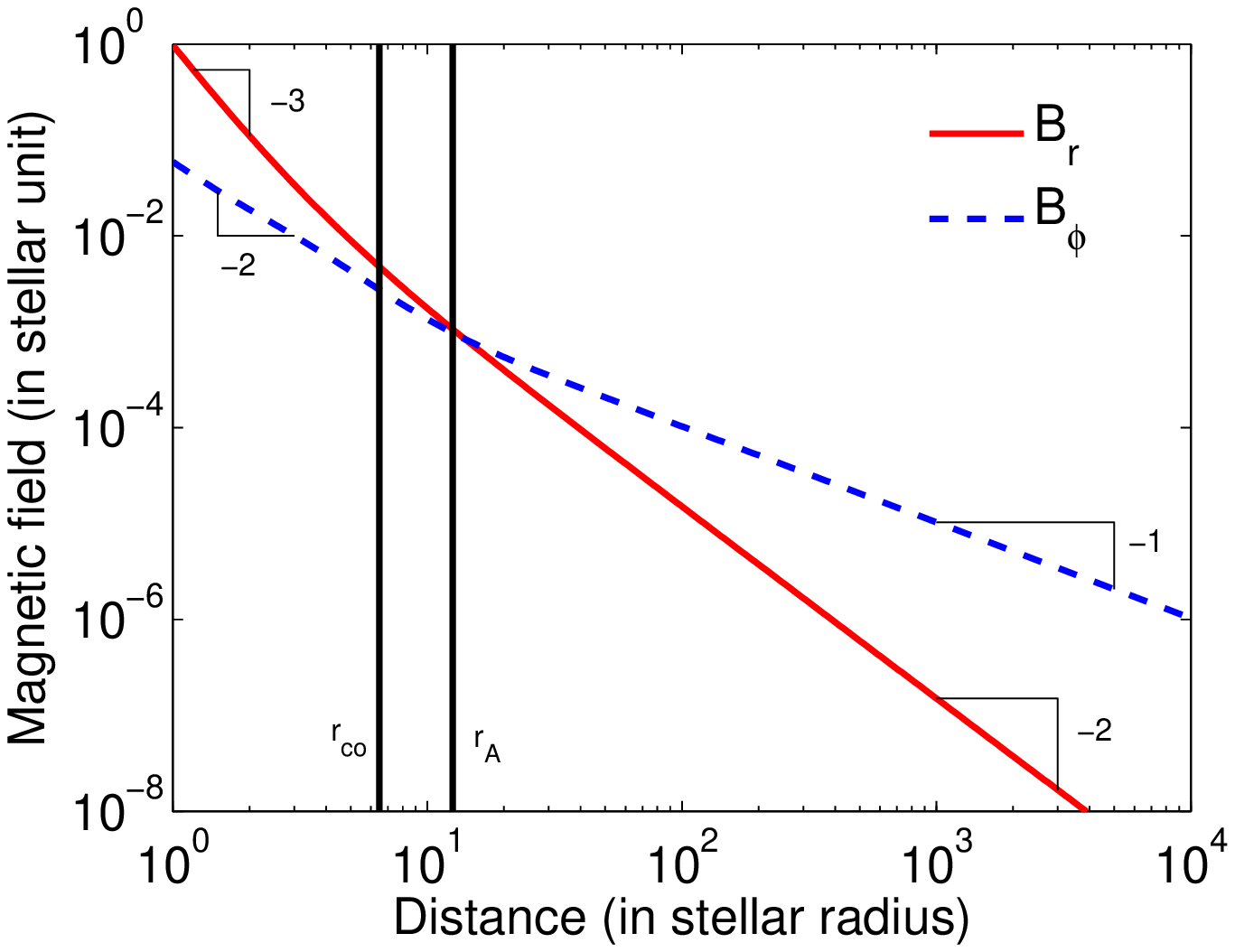}
      \caption{Radial and azimutal magnetic field of the stellar wind  for the model adapted from Zarka (2007), with the typical slopes (stellar magnetic field rescaled by its surface value of $ 1$ kG).
              }
         \label{mag}
   \end{figure}
 
\section{Results of Strugarek et al. (2014) } \label{sec:appendixC}
According to Strugarek et al. (2014) , the torque applied by the planet to the star is given by (see their Eq.~16)
\begin{equation}
\tau=K\, \tau_w\, c \left(\frac{\bpl}{B}+b \right)^p \cos^t \left( \frac{\theta_0- \Theta}{s} \right) \left(  \frac{a/\rstar}{3} \right)^{-5},
\end{equation}
where $ \tau_w=\mwind \wstar r_A^2$, where $c$, $b$, $p$, $t$, $\Theta$ and $s$ are given by their table 4, values obtained for $a/\rstar=3$, and where the last factor on the right hand side comes from their proposed scaling in $a^{-5}$ (see their Eq.~28). We have added a supplementary factor $K$ to remind that their values are all obtained for a planetary radius $\rpl/\rstar=0.1$, and the extrapolation to other planetary radius requires thus a correction $K$. We propose for instance $K=(\rpl/\rstar)^2/0.1^2$, since eq. \ref{eq:Torq} gives a torque in $R_{obs} ^2$, and eq. \ref{eq:RobsM0} gives $R_{obs} \sim \rpl$.

\section*{Acknowledgments}

We warmly thank Caroline Terquem for her input regarding tidal interactions and for a critical reading of a first version of the manuscript. It is a pleasure to acknowledge a number of enligthing discussions with P. Zarka, S. Brun, A. Strugarek, S. Matt, A. Vidotto, C. Zanni, J. Ferreira on the complex issue of star-planet-disk-wind magnetic interactions.  We would like to dedicate this short contribution to the memory of Jean-Paul Zahn, a pionier in the development of the theory of stellar tides, who passed away a week before this paper was accepted. 

\bibliographystyle{mn2e}
\bibliography{biblio}

\label{lastpage}

\end{document}